\documentclass{article}
\usepackage{psfig}

\begin{document}

\title{Derivation of a non-objective Oldroyd model from the
Boltzmann equation} 

\author{Alexander J. Wagner\thanks{awagner@ph.ed.ac.uk}\\
Department of Physics and Astronomy, University of Edinburgh,\\
                            JCMB Kings Buildings, Mayfield Road,
                            Edinburgh EH9 3JZ, U.K.
} 

\maketitle

\begin{abstract}
I calculate the hydrodynamic limit of the BGK approximation of the
Boltzmann equation for the case of a long stress relaxation time and
find that the stress obeys a viscoelastic constitutive equation. The
constitutive equation is different from standard constitutive
equations used for polymeric liquids in that it is not ``objective''
because inertial effects are important. I calculate the exact
solution for the stresses for simple shear flow and elongational flow.
\end{abstract}

\section{Introduction}
It has long been known that not only complex fluids like polymeric
liquids show viscoelastic behavior but that even gases show some
viscoelasticity \cite{maxwell}. Because these effects are small for
gases and difficult to measure in most cases they have only received a
more detailed consideration when molecular dynamics simulations showed
non-Newtonian behavior in simple shear flow \cite{naitoh} and Zwanzig
\cite{zwanzig} found an analytical solution for the BGK approximation
of the Boltzmann equation that reproduced the shear thinning found in
the computer simulations.  Examination of the non-Newtonian aspects of
flows described by the Boltzmann equation focuses on the case of a
simple shear flow \cite{dufty,gomez,lutsko,lee} and it was analyzed in
increasing detail by generalizing the results to hard spheres
\cite{gomez,lutsko}. Recently the results have been generalized to a
perturbative expansion around the simple shear flow by Lee {\it et
al.} \cite{lee}.

Giraud and d'Humi\'{e}res\cite{laurent} take a different angle on the
non-Newtonian limit of the Boltzmann equation. They have attempted
to use the viscoelastic properties hidden in the BGK-Boltzmann
equation to define lattice Boltzmann models with viscoelastic
properties. So far, however, their results are limited to linear
viscoelasticity. We hope that the analytical methods derived in this
article can help to direct the research into viscoelastic lattice
Boltzmann methods.

In this article we present an intriguingly simple analysis of the
moments of the Boltzmann equation in the BGK approximation which
allows us to derive a viscoelastic constitutive equation. This
constitutive equation has a number of interesting properties. Firstly it
is not ``objective'', {\it i.e.} inertial effects can not be
neglected. Secondly the first normal stress difference has
a negative sign as compared with a positive sign for polymeric
liquids. Unlike the upper-convected Maxwell model used to describe
polymeric liquids our constitutive equation shows shear thinning. 

In the second part of this article we will show that our constitutive
equation reduces to the analytical results derived by
Zwanzig\cite{zwanzig} for the case of simple shear flow. For this
simple situation we then give an intuitive explanation for the
difference in viscoelastic behavior for a gas and a polymeric
liquid. We also find a new exact solution for the stress for
elongational flows.

\section{Derivation of viscoelasticity in the BGK Boltzmann equation}
\label{analytic} 
The Boltzmann equation was derived for rarefied gases and is a complicated
integral equation. However in the hydrodynamic limit the
Navier-Stokes equations for fluid flow can be recovered and this is
one of the reasons that it is also the basis for a popular method of
simulating fluids, the lattice Boltzmann method. All the properties
required for the description of a fluid are retained in the somewhat
simpler single relaxation time approximation introduced by Bhatnagar
{\it et al.} \cite{Bhatnagar} (BGK) which is still a complicated non-linear
equation. In certain simple situations, however, this equation has an
analytical solution \cite{zwanzig}.

In this section we derive a viscoelastic limit of the BGK
approximation\cite{Bhatnagar} of the Boltzmann equation in the limit
of large relaxation times.  The Boltzmann equation is an evolution
equation for probability density $f_v({\bf x},t)$ to find a particle
at time $t$, at position $\bf x$ with velocity $\bf v$. The evolution
equation consists of a free streaming of the particles and a collision
term. The effect of the collision term is approximated in the BGK
approximation as a simple relaxation of the distribution function
towards the equilibrium distribution $f^0$.  This local equilibrium
distribution is the Maxwell-Boltzmann distribution for a given
density, velocity and temperature. The evolution equation is
\begin{equation} \label{eqn1}
\partial_t f_v + v_\alpha \partial_\alpha f_v = \frac{1}{\tau} (f^0_v -f_v)
\end{equation}
where the relaxation time $\tau$ is $\tau(\rho,\theta)= \tau_0/(\rho
T^b)$. The exponent $b$ can be calculated for molecules with potential
$V(r)\sim r^{-l}$ to be $b=1/2-2/l$ \cite{zwanzig}. The simplest case is the case of
so called Maxwell molecules ($l=4$) for which $b=0$ and there is not
$T$-dependence.  We define the density
$\rho$, the mean velocity $u_\alpha$ and the temperature $\theta$ as
\begin{equation}
\int f_v dv=\rho,\;\;\; \int f_v v_\alpha = \rho u_\alpha,
\;\;\; \int f_v ({\bf v-u})^2 = D \rho \theta
\end{equation}
where $D$ is the number of spatial dimensions.  Because of mass,
momentum and energy conservation the equilibrium distribution has to
have the same moments 
\begin{equation}
\int f^0_v dv=\rho,\;\;\; \int f^0_v v_\alpha = \rho u_\alpha,
\;\;\; \int f^0_v (v_\alpha-u_\alpha)(v_\beta-u_\beta) = \rho
\theta\delta_{\alpha\beta} 
\end{equation}
and the second moment is isotropic.  We can now integrate the
Boltzmann equation over the conserved moments $1, {\bf v}, {\bf v}^2$
corresponding to mass, momentum and energy and get the continuity
equation
\begin{equation}
\partial_t \rho + \partial_\alpha (\rho u_\alpha) =0
\end{equation}
and a momentum conservation equation
\begin{equation}
\rho \partial_t u_\alpha +\rho
 u_\beta \partial_\beta u_\alpha =- \partial_\beta \Pi_{\alpha\beta}
\end{equation}
where
\begin{equation}
\Pi_{\alpha\beta}= \int f_v (v_\alpha-u_\alpha) (v_\beta-u_\beta) d{\bf v}
\end{equation}
We now need to calculate $\Pi_{\alpha\beta}$ and we split the
contribution to $\Pi_{\alpha\beta}$ into an equilibrium and a traceless
non-equilibrium part where we use 
\begin{equation}\label{approx}
f_v=f_v^0-\tau (\partial_t f_v +
v_\alpha \partial_\alpha f_v)
\end{equation}
 from equation (\ref{eqn1}).
\begin{eqnarray}
\Pi_{\alpha\beta} &=& \int f_v (v_\alpha-u_\alpha) (v_\beta-u_\beta) d{\bf v}\nonumber\\
&=& \int \left[f^0_v - \tau (\partial_t f_v + \partial_\gamma v_\gamma
f_v)\right ] (v_\alpha-u_\alpha) (v_\beta -u_\beta)\nonumber\\
&=& \rho\theta\delta_{\alpha\beta} + \sigma_{\alpha\beta} \label{Peqn}
\end{eqnarray}
where $\sigma_{\alpha\beta}$ is a traceless stress.

The energy conservation equation is then
\begin{eqnarray}
0 &=&\int (\partial_t f_v v_\alpha v_\alpha+ v_\beta \partial_\beta f_v
v_\alpha v_\alpha) dv\nonumber\\
&=&\partial_t (\Pi_{\alpha\alpha}+\rho u_\alpha u_\alpha)
\nonumber\\
&&+ \partial_\beta (Q_{\beta\alpha\alpha}
+\Pi_{\beta\alpha}u_\alpha+
\Pi_{\alpha\beta}u_\alpha+\Pi_{\alpha\alpha} u_\beta
+\rho u_\beta u_\alpha u_\alpha)\label{energy}
\end{eqnarray}
where $Q$ is related to the third velocity moment of the distribution
function
\begin{eqnarray}
Q_{\alpha\beta\gamma}&=&\int f_v (v_\alpha-u_\alpha) (v_\beta-u_\beta)
 (v_\gamma-u_\gamma) d{\bf v} \label{Qeqn}\\
&=& \int f_v v_\alpha v_\beta v_\gamma dv -u_\alpha \Pi_{\beta\gamma}
 -u_\beta \Pi_{\alpha\gamma} -u_\gamma \Pi_{\alpha\beta}-\rho
 u_\alpha u_\beta u_\gamma.\nonumber
\end{eqnarray}
We can restate the energy conservation equation (\ref{energy}) as
\begin{eqnarray}
\partial_t (\rho \theta) + u_\beta \partial_\beta(\rho\theta) 
&=& -\frac{2+D}{D}
\rho\theta \partial_\beta u_\beta- \frac{1}{D} \partial_\beta
Q_{\beta\alpha\alpha}
 - \frac{2}{D} \partial_\beta u_\alpha \sigma_{\alpha\beta}\label{teqn}
\end{eqnarray}
 The equation (\ref{Peqn}) imposes
a constitutive equation for the stress $\sigma_{\alpha\beta}$:
\begin{eqnarray}
\frac{1}{\tau}\sigma_{\alpha\beta}&=&
-\int (\partial_t f_v+ v_\gamma \partial_\gamma f_v)
(v_\alpha-u_\alpha) (v_\beta -u_\beta)\nonumber\\
&=&-\partial_t (\rho\theta\delta_{\alpha\beta}+\sigma_{\alpha\beta})
- \partial_\gamma u_\beta (\rho\theta\delta_{\alpha\gamma}+\sigma_{\alpha\gamma})
\nonumber \\
&&-\partial_\gamma u_\alpha (\rho\theta\delta_{\gamma\beta}+\sigma_{\gamma\beta})
-\partial_\gamma u_\gamma (\rho\theta\delta_{\alpha\beta}+\sigma_{\alpha\beta})
\nonumber \\
&&-u_\gamma \partial_\gamma (\rho\theta\delta_{\alpha\beta}+\sigma_{\alpha\beta})
-\partial_\gamma Q_{\alpha\beta\gamma}
\end{eqnarray}
Using the
equation for the energy conservation we get
for the $\theta$ terms
\begin{eqnarray}
&& \partial_t (\rho\theta) \delta_{\alpha\beta}
+ u_\gamma \partial_\gamma (\rho\theta)\delta_{\alpha\beta} +
\rho\theta (\partial_\beta u_\alpha+\partial_\alpha u_\beta 
+\partial_\gamma u_\gamma \delta_{\alpha\beta})\nonumber \\
&=&\rho\theta(\partial_\beta u_\alpha+ \partial_\alpha u_\beta
-\frac{2}{D} \partial_\gamma u_\gamma\delta_{\alpha\beta})
+ \frac{2}{D} \partial_\gamma u_\delta \sigma_{\gamma\delta}
-\frac{1}{D} \partial_\gamma Q_{\gamma\delta\delta}.
\end{eqnarray}
Introducing a convected time derivative as 
\begin{equation}
\sigma_{\alpha\beta<1>}=D_t \sigma_{\alpha\beta}+
(\partial_\gamma u_\alpha \sigma_{\gamma\beta} 
+\sigma_{\alpha\gamma} \partial_\gamma u_\beta),
 \label{tderiv}
\end{equation}
where the total derivative is $D_t=\partial_t + v_\alpha
\partial_\alpha$, we can now write the constitutive equation as
\begin{eqnarray}
&&\sigma_{\alpha\beta}+\tau \sigma_{\alpha\beta <1>}
- \tau\frac{2}{D} (\sigma_{\gamma\delta}\partial_\delta u_\gamma) 
\delta_{\alpha\beta}
+ \tau \partial_\gamma u_\gamma \sigma_{\alpha\beta}
\nonumber\\
&=&-\tau\rho\theta
(\partial_\alpha u_\beta+ \partial_\beta u_\alpha -\frac{2}{D}
\partial_\gamma u_\gamma \delta_{\alpha\beta})
-\tau \partial_\gamma Q_{\alpha\beta\gamma} +\tau \frac{1}{D} \partial_\gamma
Q_{\gamma\delta\delta}\delta_{\alpha\beta} \label{const_p}
\end{eqnarray}
This expression is similar to the stress relaxation equation proposed
by Maxwell\cite{maxwell} and we get his result if we replace the
convected derivative with a partial time
derivative. Eqn. (\ref{const_p}) was first derived in its complete
form from an expansion of the full Boltzmann equation in Hermite
polynomials by Grad\cite{grad}.

The usual expansion of the Boltzmann equation in the hydrodynamic
limit is done under the assumption $\tau = O(1)$ and $\partial =
O(\epsilon)$. Then we get for the leading order of the stress (of
order $\epsilon$)
\begin{equation}
\sigma =- \tau \rho\theta \left(\nabla u + (\nabla u)^\dagger - \frac{2}{D}
tr(\nabla u)\delta\right) \label{sigeqn}
\end{equation}
For larger relaxation times, however, we obtain a viscoelastic
result. We keep the assumption that derivatives are small to order
epsilon $\partial = O(\epsilon)$ but keep terms up to order
$O(\epsilon^2)$.  We now get to leading order $O(\epsilon^2)$ the full
equation (\ref{const_p}) but the terms containing $Q$ still need to be
expressed in terms of macroscopic quantities.  The contribution of the
equilibrium distribution to $Q$ vanishes so it is useful to express
$Q$ in terms of the first order non-equilibrium contributions. If we
iteratively substitute eqn. (\ref{approx}) twice into eqn.(\ref{Qeqn})
we can write
\begin{eqnarray}
Q_{\alpha\beta\gamma}&=& 
- \tau \left(\int \partial_t f^0_v 
(v_\alpha-u_\alpha)(v_\beta-u_\beta)(v_\gamma-u_\gamma)
\right.\nonumber\\&&\left.
+\int \partial_\delta f^0_i v_\delta
(v_\alpha-u_\alpha)(v_\beta-u_\beta)(v_\gamma-u_\gamma)
\right)+O(\partial^2)
\nonumber\\
&=& \tau \left\{ \int f^0_i \partial_t 
\left[ (v_\alpha-u_\alpha)(v_\beta-u_\beta)(v_\gamma-u_\gamma)\right]
\right. \nonumber\\&&\left.
+\int f^0_i v_\delta \partial_\delta
\left[ (v_\alpha-u_\alpha)(v_\beta-u_\beta)(v_\gamma-u_\gamma)\right]
\right.\nonumber\\&&\left.
-\partial_\delta \int f^0_v 
(v_\alpha-u_\alpha)(v_\beta-u_\beta)(v_\gamma-u_\gamma)(v_\delta-u_\delta)
\right\}
+O(\partial^2)\nonumber\\
&=&-\tau \left\{(\delta_{\alpha\beta}\delta_{\gamma\delta}
+\delta_{\alpha\gamma}\delta_{\beta\delta}
+\delta_{\alpha\delta}\delta_{\beta\gamma})
\partial_\delta(\rho\theta^2)
\right.\nonumber\\&&\left.
+(\partial_t u_\alpha + u_\delta \partial_\delta u_\alpha)
\rho\theta\delta_{\beta\gamma}
+(\partial_t u_\beta + u_\delta \partial_\delta u_\beta)
\rho\theta\delta_{\alpha\gamma}
\nonumber\right.\\&&\left.
+(\partial_t u_\gamma+u_\delta \partial_\delta u_\gamma)
\rho\theta\delta_{\alpha\beta}\right\}+O(\partial^2)
\nonumber\\
&=&
-\tau \left\{(\delta_{\alpha\beta}\delta_{\gamma\delta}
+\delta_{\alpha\gamma}\delta_{\beta\delta}
+\delta_{\alpha\delta}\delta_{\beta\gamma})
\partial_\delta(\rho\theta^2)
\right.\nonumber\\&&\left.
- (\delta_{\alpha\beta}\delta_{\gamma\delta}
+\delta_{\alpha\gamma}\delta_{\beta\delta}
+\delta_{\alpha\delta}\delta_{\beta\gamma})
\theta\partial_\delta(\rho\theta)
\right\}+O(\partial^2)\nonumber\\
&=& -\tau\rho\theta(\delta_{\alpha\beta}\delta_{\gamma\delta}
+\delta_{\alpha\gamma}\delta_{\beta\delta}
+\delta_{\alpha\delta}\delta_{\beta\gamma})
\partial_\delta\theta +O(\partial^2)
\label{refQ}
\end{eqnarray}
which is the only approximation we have to make.  We then get for the $Q$
terms in eqn. (\ref{sigeqn})
\begin{eqnarray}
&&\partial_\gamma Q_{\alpha\beta\gamma} -\frac{1}{D} 
\partial_\gamma Q_{\gamma\delta\delta} \delta_{\alpha\beta}\nonumber\\
&=&
-\partial_\alpha(\tau\rho\theta \partial_\beta \theta)
-\partial_\beta(\tau\rho\theta \partial_\alpha\theta)
+\frac{2}{D}\partial_\gamma(\tau\rho\theta \partial_\gamma\theta)
\delta_{\alpha\beta}
+O(\epsilon^2)
\nonumber\\
&=&
-\tau\rho\theta \left(2\partial_\alpha\partial_\beta
-\frac{2}{D}\partial_\gamma\partial_\gamma\delta_{\alpha\beta}\right)
+O(\epsilon^2)
\end{eqnarray}
where we have used $\tau(\rho\theta)=\tau_0/(\rho\theta)$ in the last line.
 So the viscoelastic constitutive equation is
\begin{eqnarray}
\sigma_{\alpha\beta}+\tau \sigma_{\alpha\beta <1>}&=&\tau\rho\theta
(\partial_\alpha u_\beta+ \partial_\beta u_\alpha -\frac{2}{D}
\partial_\gamma u_\gamma \delta_{\alpha\beta})
\nonumber \\&&
-\tau\rho\theta \left(2\partial_\alpha 
\partial_\beta- \frac{2}{D} \partial_\gamma \partial_\gamma\right) 
\theta + O(\epsilon^3)
 \label{const}
\end{eqnarray}
where the last term in $\theta$ represents stresses induced by second
derivatives of the temperature. Note that this constitutive equation
is exact if second order derivatives vanish.

\section{Comparison to usual ``objective'' constitutive equations}
In the theory of viscoelastic constitutive equations usually the
assumption is made that the constitutive equation should be invariant
under an arbitrary transformation that preserves distances and time
intervals \cite{bird}. That also requires form-invariance of the
constitutive equation under some non-inertial transformations such as
rotations and accelerated systems. The underlying idea is that the
stresses in a polymeric fluid only the stretching of the system is
relevant and all inertial contributions can be neglected. In
particular a convected time derivative that obeys this symmetry is
said to be an ``objective time derivative''.

There are two such derivatives; the upper convected derivative
\begin{equation}
\sigma_{\alpha\beta(1)}=D_t \sigma_{\alpha\beta}-
(\partial_\gamma u_\alpha\sigma_{\gamma\beta}
+\sigma_{\alpha\gamma}\partial_\gamma u_\beta)\label{upper}
\end{equation}
and the lower convected derivative
\begin{equation}
\sigma_{\alpha\beta}^{(1)}=D_t \sigma_{\alpha\beta}+
(\partial_\alpha u_\gamma \sigma_{\gamma\beta}
+\sigma_{\alpha\gamma}\partial_\beta u_\gamma).\label{lower}
\end{equation}
Any linear combination of these two is also objective, and in the past
the Jaumann derivative given by
$(\sigma_{\alpha\beta(1)}+\sigma_{\alpha\beta}^{(1)})/2$ has been
popular. Today, however, constitutive equations for viscoelastic
fluids are constructed with a broad preference for the upper convected
derivative because these agree best with the new
experimental results \cite{bird}.

Many frequently used constitutive models are contained in the rather
general Oldroyd 8-constant model \cite[7.3-2]{bird}. If we only keep
constants that can be related to our constitutive equation we get
\begin{equation}
\sigma_{\alpha\beta} + \lambda_1 \sigma_{\alpha\beta(1)} + 
\frac{1}{2} \lambda_6 \partial_\gamma u_\delta \sigma_{\delta\gamma}
\delta_{\alpha\beta} = -\eta_0 (\partial_\alpha u_\beta +
\partial_\beta u_\alpha)
\end{equation}
as a special case of the Oldroyd 8-constant model. This is identical
to our constitutive model of eqn.(\ref{const}) up to the sign change
between $\sigma_{\alpha\beta(1)}$ and $\sigma_{\alpha\beta<1>}$ if we
consider that the Oldroyd model is only derived for incompressible
fluids ($\partial_\gamma u_\gamma = 0$) and does not consider elastic
effects of temperature gradients ($\partial \theta = 0$). It is
interesting to note that the most frequently used special cases of the
Oldroyd 8-constant model all assume that $\lambda_6 =0$. (The equation
above with $\lambda_6 = 0$ is known as the upper convected Maxwell
model).

The difference between the Oldroyd model and our constitutive equation
lies purely in the nature of the convected derivatives $\sigma_{(1)}$
of eqn. (\ref{upper}) and $\sigma_{<1>}$ of eqn. (\ref{tderiv}) which
is mathematically a difference in the sign of the $\partial u \sigma$
terms. In particular this means that the new convected derivative is
not ``objective''.  It is not too surprising, however, that inertial
effects are important for a gas and therefore there is no reason to
require an objective derivative for the viscoelastic constitutive
equation of a gas.  We will be discussing the effect of this subtle
difference for two special cases, the simple shear flow and the
elongational flow and give also an intuitive explanation for the
differences in the next section.

\subsection{Simple shear flow}
The most studied situation is probably the
simple shear flow with velocity profile 
\begin{equation}
{\bf u}=\left(\begin{array}{c}\dot{\gamma}y\\0\\0\end{array} \right) 
\label{eqnu}
\end{equation}
where $\dot{\gamma}$ is the shear rate.
The stress is then given by
\begin{equation}
\sigma = \left(
\begin{array}{ccc}
\sigma_{xx} & \sigma_{xy} & 0\\
\sigma_{xy} & \sigma_{yy} & 0\\
0 & 0& \sigma_{zz}
\end{array} \right)
\end{equation}
and we obtain the differential equations for the stress terms
\begin{eqnarray}
\label{str1eqn}
\sigma_{xx} + \tau \partial_t \sigma_{xx} &=& \tau \left(\frac{2}{3}-2\right)
\dot{\gamma} \sigma_{xy}\\
\sigma_{yy} + \tau \partial_t  \sigma_{yy} &=& \tau \frac{2}{3} 
\dot{\gamma} \sigma_{xy}\\
\sigma_{xy} + \tau \partial_t \sigma_{xy} + \tau \dot{\gamma}
\sigma_{yy} &=& - \tau \rho \theta \dot{\gamma} 
\end{eqnarray}
To close this system of equations we also need the heat equation
(\ref{teqn}) which reads for a spatially constant temperature and a
divergence free velocity field
\begin{equation}
\partial_t (\rho \theta)=-\frac{2}{D}\partial_\alpha u_\beta \sigma_{\beta\alpha}
\label{strteqn}
\end{equation}
We now make the Ansatz
\begin{equation}
\sigma_{\alpha\beta}(\theta)=\rho\theta \hat{\sigma}_{\alpha\beta}
\end{equation}
so that we get from (\ref{strteqn}) for the time dependence of the temperature
\begin{equation}
\theta(t) = \theta_0 \exp\left(-\frac{2}{D}
\partial_\alpha u_\beta \hat{\sigma}_{\beta\alpha}t
\right) \label{eqntemp}
\end{equation}
which reads for the velocity field of eqn. (\ref{eqnu})
\begin{equation}
\theta(t) = \theta_0 \exp\left(-\frac{2}{3}
\dot{\gamma} \hat{\sigma}_{xy}t
\right).
\end{equation}
So the differential equations for $\sigma$ become algebraic equations
\begin{eqnarray}
\sigma_{xx} + \tau \left(-\frac{2}{3} \dot{\gamma}\hat{\sigma}_{xy}\right)
\sigma_{xx} &=& \tau \left(\frac{2}{3}-2\right)
\dot{\gamma} \sigma_{xy}\\
\sigma_{yy} + \tau \left(-\frac{2}{3} \dot{\gamma}\hat{\sigma}_{xy}\right)
  \sigma_{yy} &=& \tau \frac{2}{3} 
\dot{\gamma} \sigma_{xy}\\
\sigma_{xy} + \tau \left(-\frac{2}{3} \dot{\gamma}\hat{\sigma}_{xy}\right)
\sigma_{xy} + \tau \dot{\gamma}
\sigma_{yy} &=& - \tau \rho \theta \dot{\gamma} 
\end{eqnarray}
Thus we arrive at a cubic equation for $\sigma_{xy}$ 
\begin{equation}
\left( \frac{2}{3} \,
\frac{\tau\dot{\gamma}}{\rho\theta}\sigma_{xy}-1\right)^2 \sigma_{xy} =
-\tau\rho\theta\dot{\gamma} 
\end{equation}
The solution can be written in analytic terms for $\tau=\tau_0/\rho$ as
\begin{eqnarray}
\theta(t) &=& \theta_0 \exp(t\lambda)\\
\sigma_{xx} &=& \rho \theta(t) \left(\frac{1}{1+\tau\lambda}
+\frac{2 \tau^2\dot{\gamma}^2}{(1+\tau \lambda)^3}\right)\\
\sigma_{yy} &=& \rho \theta(t) \frac{1}{1+\tau \lambda}\\
\sigma_{zz} &=& \rho \theta(t) \frac{1}{1+\tau \lambda}\\
\sigma_{xy} &=& - \frac{3}{2} \rho \theta(t) \frac{\lambda}{\dot{\gamma}} 
\end{eqnarray}
with 
\begin{equation}
\lambda = \frac{4}{3\tau} \sinh^2\left(
\frac{1}{6} \cosh^{-1}(1+9 \tau^2
\dot{\gamma}^2)\right)
\end{equation}
which is also the exact solution for the BGK approximation of the
Boltzmann equation first described by Zwanzig \cite{zwanzig} where he
derived $f_v(t)$ from eqn. (\ref{eqn1}) for a simple shear flow. He
then used this solution in the equivalent of eqn. (\ref{Peqn}) to
determine the stress. We have neglected transient terms which are
decaying exponentially. We recover the exact result because our
approximation of eqn.(\ref{refQ}) become exact for spatially constant
temperature.

There are three invariants of the stress tensor which are defined
as $\sigma_{xy}=-\eta \dot{\gamma}$ related to the viscosity $\eta$,
$\sigma_{xx}-\sigma_{yy}=-\Psi_1 \dot{\gamma}^2$ related to the first
normal stress coefficient $\Psi_1$, and $\sigma_{yy}-\sigma_{zz}=-\Psi_2
\dot{\gamma}^2$ related to the second normal stress coefficient $\Psi_2$.
For our constitutive equation we get
\begin{equation}
\eta=\frac{\tau \rho \theta}{(1+\tau\lambda)^2}
\;\;\; \Psi_1=- \frac{2\tau\eta}{1+\tau\lambda}
\;\;\; \Psi_2=0 \label{viscomeqn}
\end{equation}
In this simple situation we get a result that is equivalent to that of
an upper convected Maxwell model but with the opposite sign for the first
normal stress difference. Viscoelastic polymeric systems for which the
upper convected Maxwell model was devised seem to
always have a positive first normal stress coefficient $\Psi_1$. In the
next section we will give some intuitive understanding for this
fundamental difference in the viscoelasticity.

\begin{figure}
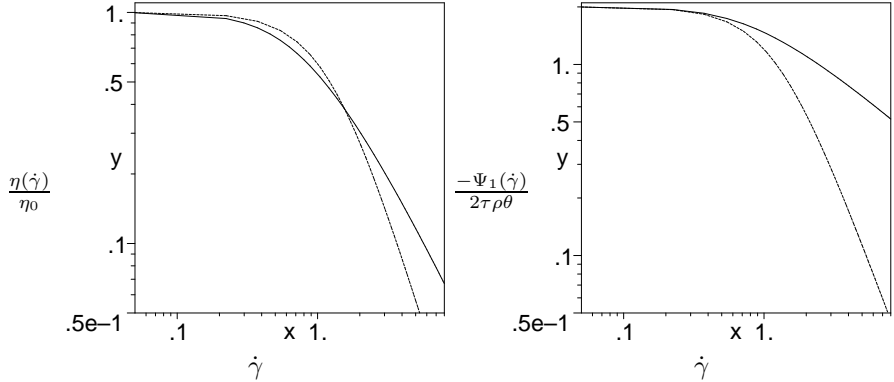

\begin{minipage}{\textwidth}
\begin{center}
\begin{minipage}{0.7cm}
$\frac{\eta(\dot{\gamma})}{\eta_0}$
\end{minipage}
\begin{minipage}{5cm}
\psfig{figure=eta.eps,height=4.5cm}
\begin{center}
$\dot{\gamma}$
\end{center}
\end{minipage}
\begin{minipage}{0.7cm}
$\frac{-\Psi_1(\dot{\gamma})}{2\tau\rho\theta}$
\end{minipage}
\begin{minipage}{5cm}
\psfig{figure=psi.eps,height=4.5cm}
\begin{center}
$\dot{\gamma}$
\end{center}
\end{minipage}
\end{center}
\end{minipage}
\caption{The exact solution for the shear-rate dependent viscosity
$\eta(\dot{\gamma})$ and first normal stress difference
$\Psi_1(\dot{\gamma})$ against the shear rate $\dot{\gamma}$ (full
line) and the isothermal solution (broken line). See text for details.}
\label{fig1}
\end{figure}

Although for a simple shear flow no stationary solution exists due to
viscous heating a stationary state can be achieved with the help of a
thermostat. The simplest way of implementing a thermostat is simply to
assume that the partial time derivatives and the
spatial derivatives of the temperature in eqns
(\ref{str1eqn})-(\ref{strteqn}) vanishes \cite{laurent,bird}. This is
equivalent to a situation where a thermostat insures a constant
temperature, but does not otherwise influence the dynamics. This
thermostat will change the energy equation but not the constitutive
equation. In this case we get
\begin{equation}
\eta=\frac{\tau \rho \theta}{1+\frac{2}{3} \tau^2 \dot{\gamma}^2}
\;\;\; \Psi_1=-2\tau\eta\;\;\; \Psi_2=0
\end{equation}
and we see that the shear-thinning persists although it now has a
different form. The first normal stress difference now only depends on
the shear rate through the viscosity. The viscosity and the first
normal stress difference for the isothermal and the heating case are
compared in Figure \ref{fig1}. Other thermostats lead to different
results. Lee and Dufty \cite{lee} considered a thermostat that is
frequently used for Molecular Dynamics simulations. This thermostat
rescales the local velocities relative to the average velocities such
that the temperature remains constant. In this case the constitutive
equation is also changed but in a way that recovers the same solution 
\ref{viscomeqn} as in the case with heating, except that the
temperature remains constant.

\subsection{Extensional flow}
We will now consider a second family of flow-fields for which, to my
knowledge, the BGK equations had not previously been solved. We will,
again, assume a spatially constant temperature so that the results
will be exact.
And extensional flow is a shear-free flow defined by the velocity profile
\begin{equation}
{\bf u} = \dot{\epsilon}\left( \begin{array}{c}
-(1+b)x/2\\
-(1-b)y/2\\
z
\end{array}
\right)
\end{equation}
where $0 \leq b \leq 1$ and $\dot{\epsilon}$ is the elongation
rate. Several special shear-free flows are obtained for particular
choices of the parameter $b$:
\begin{description}
\item{Elongational flow:} $(b=0, \dot{\epsilon}>0)$
\item{Biaxial stretching flow:} $(b=0, \dot{\epsilon}<0)$
\item{Planar elongational flow:} $(b=1)$
\end{description}
For the time-dependence of the temperature we get from
eqn. (\ref{eqntemp}) for the extensional flow
\begin{equation}
\theta(t) = \theta_0 \exp\left(-
\frac{\dot{\epsilon}}{3} (-(1+b) \hat{\sigma}_{xx}-(1-b)
\hat{\sigma}_{yy}+ 2 \sigma_{zz}) t
\right).
\end{equation}

The flow is a linear flow and if we assume a constant initial
temperature the solutions of the constitutive equation are again exact.
For the constitutive equation for the stress (\ref{const}),
using that $\sigma_{xx}+\sigma_{yy}+\sigma_{zz}=0$, we get
\begin{eqnarray}
(\frac{1}{\tau\dot{\epsilon}}-(1+b))\hat{\sigma}_{xx}
+\left((1+\frac{b}{3})\hat{\sigma}_{xx}
+(1-\frac{b}{3})\hat{\sigma}_{yy}
\right)
\left(1+\hat{\sigma}_{xx}\right)
&=& 1+b\\
(\frac{1}{\tau\dot{\epsilon}}-(1-b))\hat{\sigma}_{yy}
+\left((1+\frac{b}{3})\hat{\sigma}_{xx}
+(1-\frac{b}{3})\hat{\sigma}_{yy}
\right)
\left(1+\hat{\sigma}_{yy}\right)
&=& 1-b
\end{eqnarray}
These equations have analytic solutions one of which is
physical. These results, however, are too lengthy
to be presented here.
For the simpler case of $b=0$ the $x$ and $y$ directions are symmetric
and so we can impose $\sigma_{xx}=\sigma_{yy}$ and the equations can
be simplified to
\begin{equation}
(\frac{1}{\tau\dot{\epsilon}}-1)\hat{\sigma}_{xx}
+2 \hat{\sigma}_{xx} \left(1+\hat{\sigma}_{xx}\right)
= 1
\end{equation}
This equation has two solutions but one of these
solutions is unphysical as the stress is larger that $\rho\theta$
which would require a negative probability density $f_v$ in (\ref{Peqn}).
The physical solution is
\begin{equation}
\sigma_{xx}=-\frac{\rho\theta}{4} \frac{1+\tau\dot{\epsilon}
-\sqrt{1+2\tau\dot{\epsilon}+9\tau^2\dot{\epsilon}^2}}{\tau\dot{\epsilon}}
\end{equation}
and $-\sigma_{zz}/2=\sigma_{yy}=\sigma_{xx}$. 

\begin{figure}
\begin{minipage}{\textwidth}
\begin{center}
\begin{minipage}{0.7cm}
$\frac{\bar{\eta}_1}{\rho\theta},
 \frac{\bar{\eta}_2}{\rho\theta}$
\end{minipage}
\begin{minipage}{8cm}
\psfig{figure=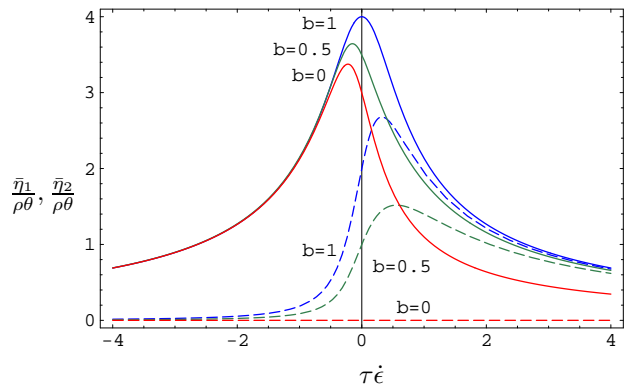,width=7.5cm}
\begin{center}
$\tau\dot{\epsilon}$
\end{center}
\end{minipage}
\end{center}
\end{minipage}
\caption{The exact solution for the extension-rate dependent
elongational viscosities $\bar{\eta}_1$(full lines) and
$\bar{\eta}_2$(dashed lines) for different parameters b (see text for
details).}
\label{figext}
\end{figure}

It is customary to define two viscosity functions to describe the
rheological behaviour of a fluid in extensional flow \cite{bird} as
\begin{eqnarray}
\sigma_{xx}-\sigma_{zz} &=& \bar{\eta}_1 (\dot{\epsilon},b)
\dot{\epsilon}\\
\sigma_{xx}-\sigma_{yy} &=& \bar{\eta}_2 (\dot{\epsilon},b) \dot{\epsilon}
\end{eqnarray}
and we show the values for different $b$ in Figure \ref{figext}. For
the elongational flow and the biaxial stretching flow ($b=0$) defined
above the $x$ and $y$-directions are equivalent and therefore
$\bar{\eta}_2 =0$. For the elongational flow ($b=0,\dot{\epsilon}>0$)
we find only shear thinning (in polymeric flows a shear-thickening is
observed) but for the biaxial stretching flow gases show a shear
thickening. 

For the planar elongational flow $b=1$ the $x$ and $z$ directions
become symmetric if you invert the extension rate
($\sigma_{xx}(\dot{\epsilon},b=1) = \sigma_{zz}(-\bar{\epsilon},
b=1)$) so that the viscosity $\bar{\eta}_1$
becomes symmetric, but the viscosity $\bar{\eta}_2$ becomes
asymmetric. Even though the planar elongational flow is a
two-dimensional flow the stress will be different for a true
two-dimensional system and this three dimensional case.

If we consider a system with a simple-minded thermostat again that
simply assumes a constant temperature we get unphysical divergences
for the stress
\begin{eqnarray}
\sigma_{xx} &=& \tau\rho\theta \dot{\epsilon}
\frac{3(1+b)+\tau\dot{\epsilon}(-3+6b+b^2)}
{-3+(3+b^2)\tau^2\dot{\epsilon}^2}
\\
\sigma_{yy} &=& \tau\rho\theta\dot{\epsilon}
\frac{3(1-b)+\tau\dot{\epsilon}(-3-6b+b^2)}
{-3+(3+b^2)\tau^2\dot{\epsilon}^2}
\\
\sigma_{zz}&=& 2\tau\rho\theta \dot{\epsilon}
\frac{-3+\tau\dot{\epsilon}(3-b^2)}
{-3+(3+b^2)\tau^2\dot{\epsilon}^2}
\end{eqnarray} 
but the quadratic behaviour for small $\dot{\epsilon}$ still agrees with
the non-isothermal case. These divergences indicate that the
simple-minded thermostat is unphysical.

It is easy to see why this unphysical behavour can occur. Because the
thermostat will regulate the temperature without influencing the
stress we can have a second moment 
\begin{equation}
\Pi_{xx} = \int f_v (v_x-u_x)^2
\end{equation}
that becomes negative which in terms reqires negative contributions to
the probability distribution. This is what makes the thermostat unphysical.

\begin{figure}
\begin{center}
\begin{minipage}{6cm}
  \begin{minipage}[b]{6cm}
    \centerline{\psfig{figure=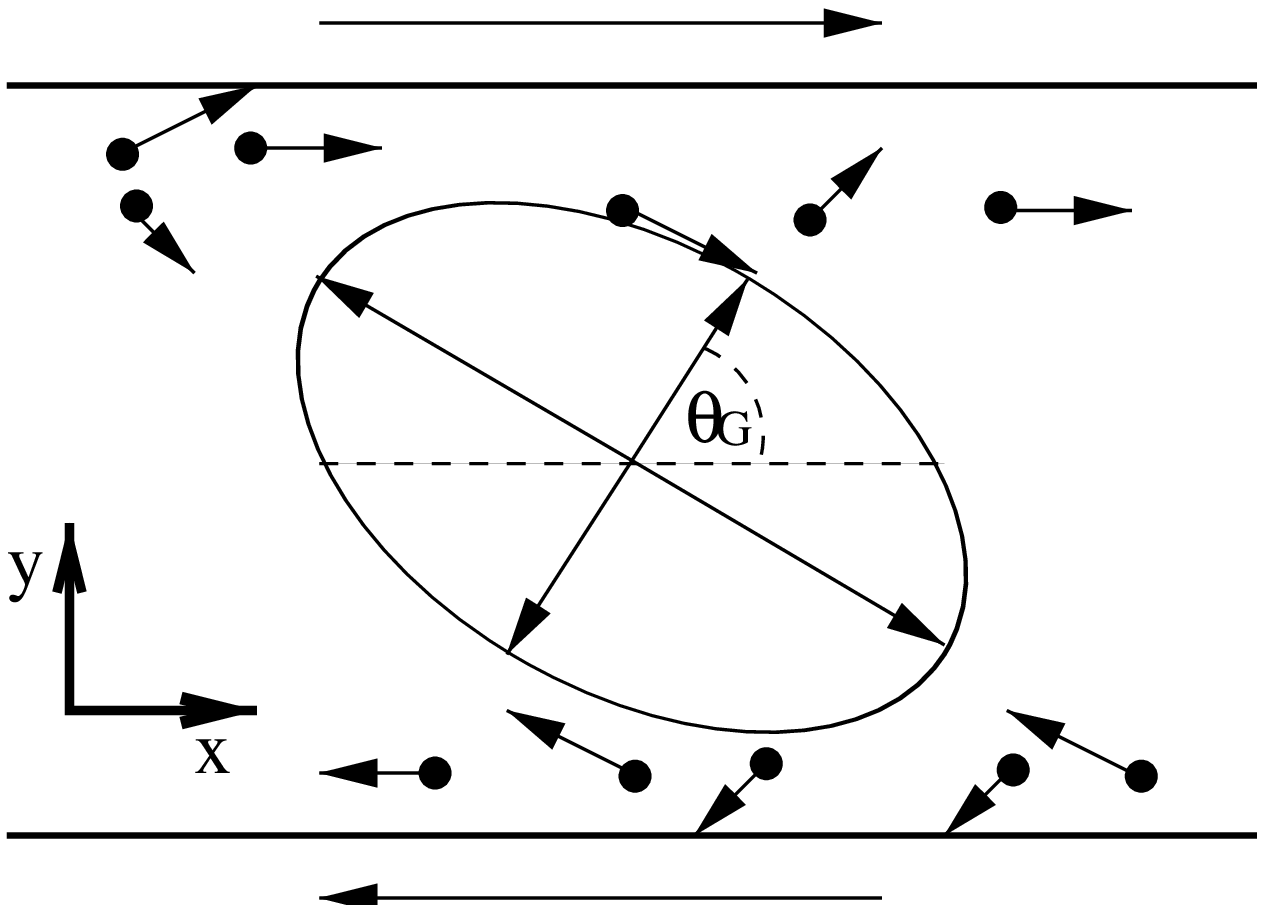,width=6cm}}
  \end{minipage}
  \begin{center}
    (a)
  \end{center}
\end{minipage}
\begin{minipage}{6cm}
  \begin{minipage}[b]{6cm}
    \centerline{\psfig{figure=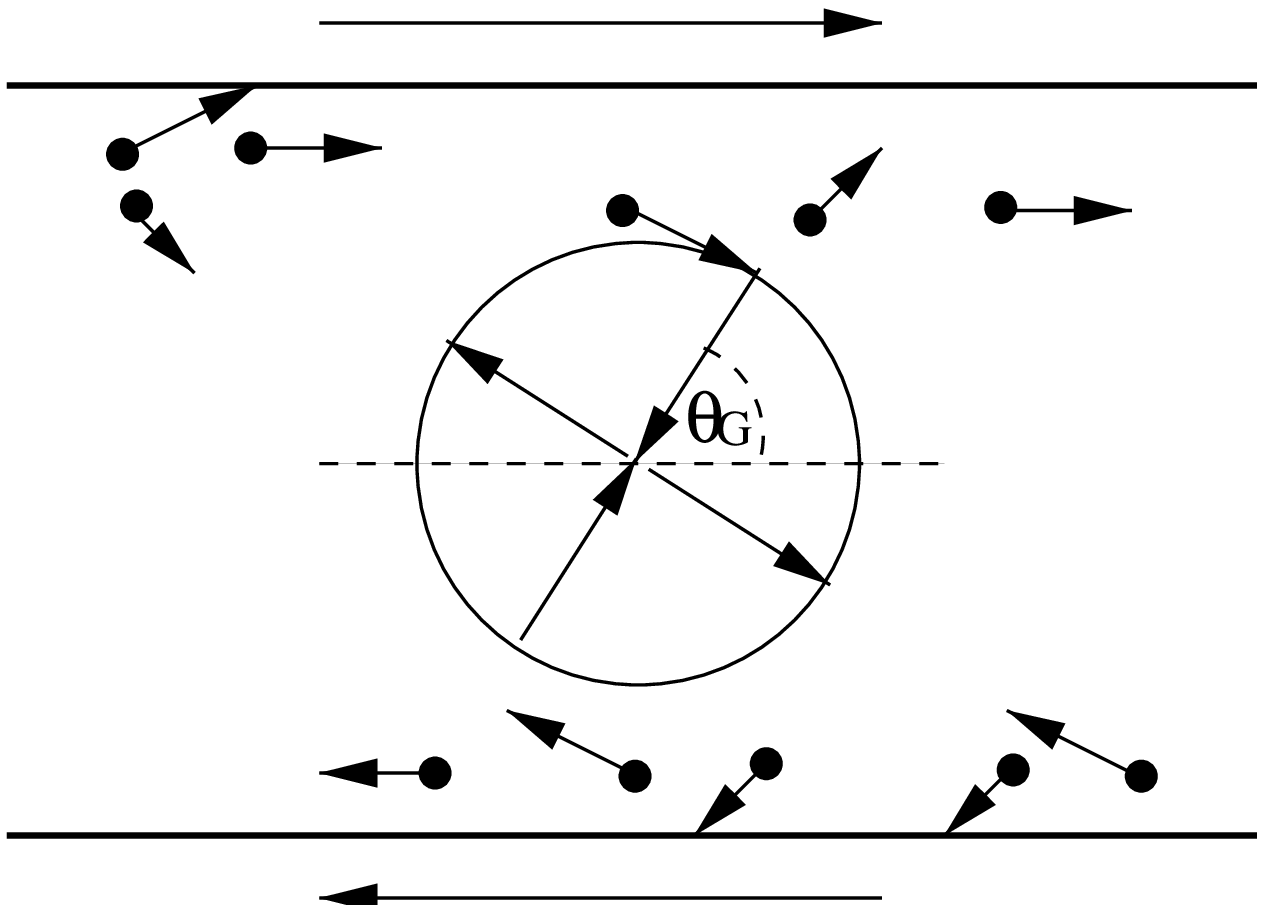,width=6cm}}
  \end{minipage}
  \begin{center}
    (b)
  \end{center}
\end{minipage}
\end{center}
\caption{Illustration of the origin of a non-diagonal pressure $\Pi$
and the corresponding stress tensor $\sigma$ in the case of a gas.
 (a) Origin of pressure tensor (b) the traceless stress tensor
(vectors pointing inwards represent a negative stress). Note that the
orientation of the stress tensor we have $\theta_g\ge 45\circ$.} 
\label{fig2}
\end{figure}

\begin{figure}
    \centerline{\psfig{figure=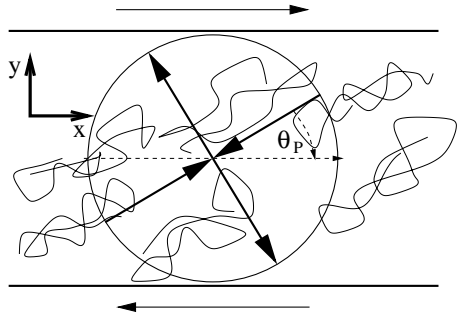,width=6cm}}
\caption{Sketch of the origin and orientation of the non-diagonal
stress tensor for a polymeric liquid in a simple shear flow. The
stress looks qualitatively similar to the gas case of Figure
\ref{fig2} but note that here $\theta_p\le 45^\circ$. Also there is
no requirement for the polymeric stress to be traceless.}
\label{fig3}
\end{figure}

\section{Intuition for the viscoelastic effects in a gas}
In Figures \ref{fig2} and \ref{fig3} we sketched the origins of the
non-isotropic stress for both the case of a gas and the case of a
polymeric liquid in a simple shear flow. In a gas the mean velocity of
the particles is given by the flow profile of equation
(\ref{eqnu}). Particles that are convected in the direction of $-y$
carry more x-momentum with them than the average at this position and
particles that are convected in the direction of $+y$ carry less
x-momentum with them that the average that the new position. Therefore
the momentum distribution will be non-isotropic as indicated in Figure
\ref{fig2}(a). If we subtract the isotropic part of the momentum
distribution we get the stress tensor $\sigma$ of equation
(\ref{Peqn}) shown in Figure \ref{fig2}(b). Note that the angle
$\theta_g$ defined in this Figure is always smaller that
$45^\circ$. Therefore the non-equilibrium stress tends to reduce the
force on the walls which is equivalent to a negative first normal
stress coefficient.

In simple shear flow the viscosity is a measure for the transport of
x-momentum in the y-direction. Because there are now fewer particles
streaming in the y-directions this also means that the viscosity is
reduced. This is the intuitive reason for the shear thinning in a gas.

In the case of a polymeric liquid the origin of the non-isotropic
stress lies in the stretching of the macromolecules as sketched in
Figure \ref{fig3}. In equilibrium without flow the macromolecules
would curl up into a spherical shape but the flow tends to deform the
shape to a more ellipsoidal from. The result is a restoring force that
would restore the molecule to a spherical shape. The first deformation
of the molecule occurs at $\theta_p=45^\circ$ and increases from
there on. The effect is that the stress distribution sketched in
Figure \ref{fig3} increases the pressure in the y-direction relative
to the pressure in the x-direction which is equivalent
to a positive first normal stress coefficient.

The deformation of the coils also means that there is less transport
of x-momentum in the y-direction which is the reason for
shear-thinning in the polymeric liquid.

\section{Conclusions}
We have derived a viscoelastic constitutive equation for a gas
described by the BGK approximation of the Boltzmann equation and shown
that this constitutive equation differs substantially from those that
describe the viscoelastic properties of polymeric materials. For
simple shear flow the exact results for the BGK approximation of the
Boltzmann equation are recovered. We explained the different sign of
the first normal stress difference in gases and polymeric liquids be
examining the qualitative different origins of the non-Newtonian stress.

\section*{Acknowledgements}
I would like to thank Gareth McKinley and Matthias Fuchs for
stimulating discussions and Mike Cates for reading the
manuscript. This work was in part funded under EPSRC Grant GR/M56234.

\def\jour#1#2#3#4{{#1} {\bf #2}, #3 (#4).}
\def\tbp#1{{\em #1}, to be published.}
\def\inpr#1{{\em #1}, in preparation.}
\def\tit#1#2#3#4#5{{#1} {\bf #2}, #3 (#4).}

\def\ap{Adv. Phys.}
\def\arf{Ann. Rev. Fluid Mech.}
\def\epl{Euro. Phys. Lett.}
\def\ijmp{Int. J. Mod. Phys. C}
\def\jcp{J. Chem. Phys.}
\def\jpc{J. Phys. C}
\def\jpcs{J. Phys. Chem. Solids}
\def\jpco{J. Phys. Cond. Mat}
\def\jsp{J. Stat. Phys.}
\def\jf{J. Fluids}
\def\jfm{J. Fluid Mech.}
\def\jnnfm{J. Non-Newtonian Fluid Mech.}
\def\pfa{Phys. Fluids A}
\def\prl{Phys. Rev. Lett.}
\def\pr{Phys. Rev.}
\def\pra{Phys. Rev. A}
\def\prb{Phys. Rev. B}
\def\pre{Phys. Rev. E}
\def\pa{Physica A}
\def\pla{Phys. Lett. A}
\def\ps{Physica Scripta}
\def\roy{Proc. Roy. Soc.}
\def\rmp{Rev. Mod. Phys.}
\def\zpb{Z. Phys. B}

\end{document}